\documentclass[conference,comsoc]{IEEEtran}
\IEEEoverridecommandlockouts
\usepackage{cite}
\usepackage{amsmath,amssymb,amsfonts}
\usepackage{algorithmic}
\usepackage{graphicx}
\usepackage{textcomp}
\usepackage{xcolor}
\usepackage{accents}
\usepackage{booktabs}

\usepackage{tikz}
\usetikzlibrary{automata, positioning, arrows}

\usepackage{subcaption}%
    
\newcommand{\intd}[1]{\;\mathrm{d}{#1}}

\usepackage[hidelinks]{hyperref}
\usepackage[nameinlink]{cleveref}
\crefname{figure}{Fig.}{Figs.}             %
\crefname{table}{Table}{Tables} %
\crefname{algorithm}{Algorithm}{Algorithms} %
\crefname{appendix}{Appendix.}{Appendices.} %
\crefformat{section}{Section~#2#1#3}

\usepackage[absolute]{textpos}\usepackage{atbegshi}\AtBeginShipout{\begin{textblock*}{.99\paperwidth}[.5,1.5](.5\paperwidth,\paperheight)\sffamily\scriptsize\noindent\textcopyright~2023 IEEE. Personal use of this material is permitted. Permission from IEEE must be obtained for all other uses, in any current or future media, including reprinting/republishing this material for advertising or promotional purposes, creating new collective works, for resale or redistribution to servers or lists, or reuse of any copyrighted component of this work in other works.\end{textblock*}}

\AtBeginShipoutFirst{%
\begin{textblock*}{.84\paperwidth}[.5,0](.5\paperwidth,.212\paperheight)
\centering\scriptsize Published in: H.-C. Lin, K.-H. Lin, and H.-Y. Wei, ``Age of Information for Power-Saving Devices with DRX Mechanism,'' in \emph{ICC 2023 - IEEE International Conference on Communications}, Rome, Italy, 2023, pp.~3234--3239, doi:~\href{https://doi.org/10.1109/ICC45041.2023.10279808}{10.1109/ICC45041.2023.10279808}.
\end{textblock*}%
}

\begin{document}

\title{Age of Information for Power-Saving Devices with DRX Mechanism}

\author{\IEEEauthorblockN{Hung-Chun Lin\IEEEauthorrefmark{1}, Kuang-Hsun Lin\IEEEauthorrefmark{1}, and Hung-Yu Wei\IEEEauthorrefmark{1}\IEEEauthorrefmark{2}}
\IEEEauthorblockA{
\IEEEauthorrefmark{1}Graduate Institute of Communication Engineering, National Taiwan University, Taipei, Taiwan. \\
\IEEEauthorrefmark{2}Department of Electrical Engineering, National Taiwan University, Taipei, Taiwan.}\thanks{Hung-Yu Wei is grateful for the funding support from Qualcomm under grant NAT-435534.}\thanks{Corresponding Author: Hung-Yu Wei (Email: hywei@ntu.edu.tw)}}
\maketitle

\begin{abstract}
Reducing the power consumption of User Equipment (UE) has always been a crucial issue in communication systems. As a result, Discontinuous reception (DRX) is introduced in today's New Radio (NR) networks as a power-saving strategy. In addition, the variety of low latency applications in 6G makes the issue of maintaining data freshness increasingly important. This paper aims to study the relationship between freshness and power-saving. To capture the information freshness, we use Age of Information (AoI) as the performance metric. To our best knowledge, unlike previous studies focusing on power saving and delay trade-offs, this is the first study considering the impact of DRX on AoI. We solve the closed-form solution of the peak average AoI with the DRX mechanism. Moreover, we propose a parameters adjustment strategy. Based on the simulation validation, our proposed strategy can enable the UE to keep the AoI close to the theoretical lower bound AoI while achieving high power-saving efficiency.
\end{abstract}


\section{Introduction}
Power saving for User Equipment (UE) is one of the main issues of wireless networks. With the advent of next-generation wireless networks offering a wide range of applications, the power requirements for devices will also escalate. Among all the applications, we notice that the requirement for information freshness has recently increased. For instance, in an intelligent factory-related application, an access point (AP) will accept the environmental data and transmit them to the other production machines in the product line. This transmission type usually requires high reliability or low latency to prevent inappropriate operations resulting from outdated information. Moreover, we expect machines to consume less energy, benefiting the environment and reducing production costs. Consequently, reducing UE power consumption while maintaining fresh data at the destination will be essential for the future network.

Discontinuous Reception (DRX) is one of the promising techniques to reduce power consumption among many power-saving mechanisms. It allows a UE to temporarily turn off the radio frequency (RF) module and not monitor its Physical DL Control Channel (PDCCH) when there is no traffic. The current 3rd Generation Partnership Project (3GPP) standard also adopts the DRX mechanism in the 5G new radio (NR) network \cite{3gpp.38.331}. However, research has shown that the DRX mechanism may deteriorate the Quality of Service (QoS) despite its power-saving advantage. Most previous studies have discussed the trade-off between power saving and packet delay\cite{Koc2014DRXmodel,Fowler2012DRXdelay,Ferng2018DRXdelay}, and a few papers have examined the relationship between the DRX mechanism and throughput\cite{Shah2021DRXthroughput}. Despite these works, we have not found any literature that discusses the impact of DRX on data freshness.

The Age of Information (AoI) has been introduced as a new performance measure to evaluate the freshness of information. It is defined as the time elapsed since the generation of a device's most recently received update packet. Early AoI research focused on the queueing theory. However, nowadays, there is a large amount of research on AoI analysis for various network services requiring timely information access, such as the Internet of Things and vehicular networks \cite{Yates2021AoIsurvey}. It is worth mentioning that delay and AoI are not equivalent. Delay is a packet-centric metric, considering the time a packet stays in the network. On the other hand, AoI describes how well the data at the destination can reflect the current status of the source at any given time and is destination-centric. 

Some studies impose a constraint on AoI to address the problem of the trade-off between power consumption and AoI. In \cite{2020ChiangBSpowsaving}, the authors study how to minimize the power consumption of base stations with cooperative information under the age constraint. The authors in \cite{Moltafet2022ageconstraint} consider the best sampling action for multiple devices, with the minimum transmission power under specific AoI constraint. In contrast, the work \cite{Wang2020ratecontrol} investigates a cross-layer scheduler that needs to select the optimal transmission rate by the system state to minimize AoI with power constraint. Finally, the works \cite{2019GuTRAQ, Gong2018retrans} discuss the transmission energy and AoI trade-off. In \cite{2019GuTRAQ}, the objective is to find the relationship between the age and energy with TARQ (truncated automatic repeat request) scheme, and the work \cite{Gong2018retrans} derives a closed-form solution for the AoI and energy consumption with retransmission times threshold. However, we observe that the term power minimization usually refers to the choice of the appropriate data transmission power rather than the power consumption at the receiver. How to reduce the power consumption at the receiver is still an open problem. 

The contributions of this work are summarized as the following:
\begin{enumerate}
    \item To our best knowledge, this is the first work considering the AoI performance under the DRX mechanism.
    \item We obtain the closed-form solution of the peak average AoI with given DRX parameters.
    \item We observe the effect of DRX parameters on the average AoI and propose a two-phase DRX service policy.
    \item The simulation results verify our proposed policy can achieve an average AoI close to the minimum theoretical AoI.
\end{enumerate}

This paper is structured as follows: we define the system model and explain the DRX mechanism in \cref{sec:system_model}. In \cref{sec:mathematics} we derive the closed-form expressions of the sleep ratio, the percentage of time where a UE turns off its RF module, and the peak average AoI. In \cref{sec:service_policy} we propose a two-phase DRX service policy. The policy provides the parameters of DRX timers to minimize average AoI under a certain sleep ratio. The results are verified through the simulations in \cref{sec:simulations}, and we conclude our work in \cref{sec:conclusion}. 
\section{System model}\label{sec:system_model}
\begin{figure}
    \centering
    \includegraphics[width=\linewidth]{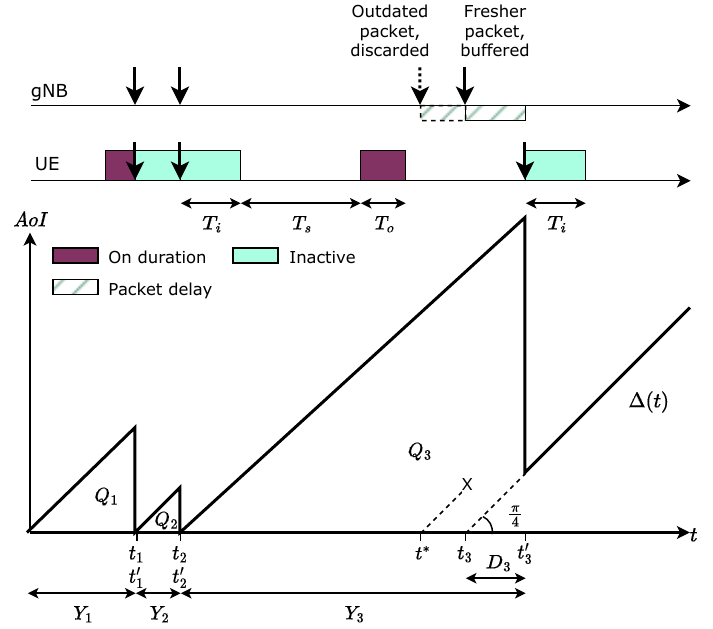}
    \caption{DRX state transition with packet arrivals, and the AoI evolution at the UE with the DRX mechanism. The arrows indicate packet arrivals. $t_1$ and $t_2$ equal to $t'_1$ and $t'_2$, respectively.}
    \label{fig:example}
\end{figure}

\begin{figure}
    \centering
    \resizebox{45ex}{18ex}{
    \begin{tikzpicture}[->,>=stealth', shorten >=1pt, auto,
    node distance=4cm, scale=0.6, 
    transform shape, align=center, line width=0.6pt,
    state/.style={circle, draw, minimum size=2cm}]
    \node[state, ] (S1) {On \\$S_1$};
    \node[state, right of=S1,] (S2) {Inactive \\$S_2$};
    \node[state, right of=S2,] (S3) {DRX sleep \\$S_3$};
    
    \draw (S1) edge[red, dashed, bend left=9ex, above, auto] node{$P_{13}$} (S3)
    (S1) edge[blue, bend left, above, auto] node{$P_{12}$} (S2)
    (S2) edge[blue, loop, in=180,out=220,looseness=5] node{$P_{22}$} (S2)
    (S3) edge[blue, bend left, above, swap] node{$P_{32}$} (S2)
    (S2) edge[red, dashed, bend left, below, swap] node{$P_{23}$} (S3)
    (S3) edge[red, dashed, bend left=9ex, below, auto] node{$P_{31}$} (S1);
    \end{tikzpicture}
    }
    \caption{DRX state transition diagram. The blue lines mean the transition with packet arrival, and the red dashed lines represent the transition with no packet arrival.}
    \label{fig:state_transition_diagram}
\end{figure}
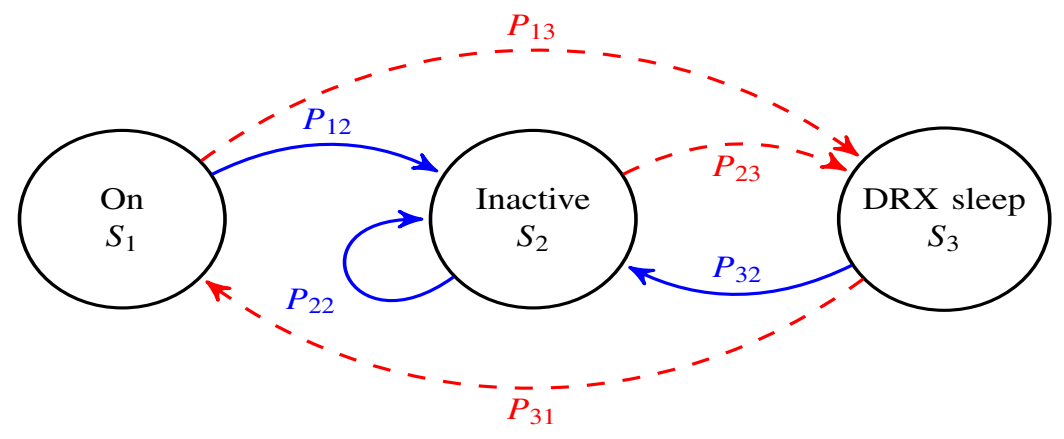

\subsection{Status update system with the DRX mechanism}
We consider a status update system with the DRX mechanism. The status updates (e.g. temperature, locations or velocities of vehicles) arrive to the Next Generation NodeB (gNB) as a rate $\lambda$ Poisson process. The generation time of status updates are same as the time they arrive at the gNB. We will use the terms packet and status update interchangeably in the rest of the paper. As with many DRX-related studies, we assume the packet transmission time from a gNB to a UE is negligible, and there is no channel error for simplicity of exposition \cite{Koc2014DRXmodel, Sharma2019DCDRX}. 

The UE is configured with the DRX mechanism. According to the 3GPP standard \cite{3gpp.38.331}, the DRX mechanism can be characterized by the on-duration timer, inactivity timer and DRX cycle. The line labeled UE in \cref{fig:example} illustrates a classic scenario of DRX operation. A UE monitors its PDCCH when the on-duration timer is running (as indicated by the purple block). If there is any downlink transmission before the on-duration timer expires, the UE will start its inactivity timer for a period $T_i$ (as indicated by the green block). Every packet arrival during the inactivity timer running will restart the inactivity timer. If there is no packet arrival before the expiry of the on-duration timer or inactivity timer, the UE will enter the DRX cycle. During the DRX cycle, the UE stops decoding PDCCH for a period of $T_s$ to save power. Then, the UE starts its on-duration timer for a period $T_o$ and monitors the downlink traffic. Since the UE will monitor the PDCCH when the on-duration timer or inactivity timer is running, we call this period active. The opposite is the dormancy period. When packet arrival occurs during the dormancy period, due to the assumption that there is no transmission time between a gNB and a UE, the UE receives the packet upon the dormancy period ends and starts its inactivity timer immediately.

In \cite{Costa2014peakAge}, the authors have verified discarding old packets waiting in the queue can significantly improve the performance of AoI. Similarly, when a UE receive buffered packets at the end of the dormancy period, only the freshest packets from past arrivals can best reduce the AoI in our system. As a result, the early generated status update can be discarded if more than one status updates arrive during the same dormancy period. The performance of the gNB is equivalent to a Preemptive Last-Generated, First-Served (P-LGFS) scheduling policy during the dormancy period. Many studies have supported that the P-LGFS policy can optimize AoI performance \cite{Yates2021AoIsurvey}.

The \cref{fig:example} shows the evolution of AoI with the DRX mechanism. The packets that can be delivered to the UE are generated at times $t_1,t_2,\cdots$, receptions at the UE occur at $t'_1, t'_2, \cdots$, and the discarded packets generate at $t^*,\cdots$. The delay of the packet $k$, denoted by $D_k$, is given by $D_k = t'_k - t_k$, and the inter-departure time $Y_k$ is defined as the time between the receptions of the $k-1$st packet and the $k$th packet, where $Y_k = t'_k - t'_{k-1}$\footnote{We read both $t_0$ and $t'_0$ as $0$.}. The function of AoI $\Delta(t)$ at a UE is defined as 
\begin{equation}
    \Delta(t) = t - u(t),%
\end{equation}
where $u(t)$ is arrival time at the gNB of UE's most recent received packet at time $t$, given by  $u(t)=\max\{t_i|t'_i \le t\}$. If a packet arrives at the gNB when the UE is active, then $t_i=t'_i$, and the age at $t'_i$ will drop to zero. This is the case for $t'_1$ and $t'_2$ in \cref{fig:example}. The packet that arrives when the UE is in the dormancy period will be discarded and not indexed if another packet comes during the same dormancy period. Otherwise, it can be received by the UE, and $t'_i$ becomes the time when the dormancy period ends. The age at $t'_i$ drops to $t'_i-t_i$ in this case, as we can see at $t'_3$. We adopt the long-term time-average AoI (TAoI) $\Delta$ to evaluate the overall AoI performance, where 
\begin{equation}\label{eq:definition_of_TAoI}
    \Delta = \lim_{T\to\infty}\frac{1}{T}\int_{0}^{T}\Delta(t)\intd{t}.
\end{equation}

\subsection{DRX states transition}
We model the DRX mechanism as a Semi-Markov process with states On, Inactive and DRX sleep. The state transition diagram is shown in \cref{fig:state_transition_diagram}, and the state transition probability from the state $S_i$ to the state $S_j$ is denoted by $P_{ij}$. The explanation for each state is as the following:
\begin{enumerate}
    \item \textbf{On $S_1$}: The UE is in this state if its on-duration timer has run, but its inactivity timer has not. When a UE enters the on state, it starts monitoring PDCCH. If there is no packet arrival during a period of $T_o$, the UE will enter DRX sleep state $S_3$. Otherwise, the UE enters state inactive $S_2$ immediately upon it receives a packet.
    \item \textbf{Inactive $S_2$}: No matter whether the on-duration timer expires or not, the UE is in this state if the inactivity timer has run. A UE in this state will also open its RF module and monitor PDCCH for a period of $T_i$. If there is packet arrival in this state, the inactivity timer will restart, and the UE needs to monitor PDCCH again for time $T_i$. Otherwise, the UE moves to $S_3$.
    \item \textbf{DRX sleep $S_3$}: The UE is in this state when it experiences the dormancy period. In this state, a UE turns off its RF module to save energy for a fixed period of $T_s$. Because the gNB knows the DRX configuration of the UE, if there is packet arrival, the gNB can transmit the freshest packet to the UE without delay when DRX sleep ends. Thus the UE will move to $S_2$ immediately if a packet is coming during DRX sleep. Otherwise, the UE will enter $S_1$.
\end{enumerate}

The gNB can configure the on-duration timer $T_o$, inactivity timer $T_i$ and sleep time $T_s$ for the UE. In addition, the sleep time should have a minimum value $T_{s}^{\min}$ to prevent a UE from switching its state too frequently, which does not match the behavior of the UE in the real world.

\section{Sleep ratio and Average age with DRX mechanism}\label{sec:mathematics}
We use the sleep ratio, the percentage of time a UE stays in DRX sleep, as the metric for power saving. For the information freshness, we give the closed-form solution of the peak average Age of Information (PAoI).
\subsection{Sleep ratio $\gamma$}
We first obtain the sleep ratio of the system by the following analysis. According to the state transition diagram in fig. \ref{fig:state_transition_diagram}, we can derive the state transition probabilities for the semi-Markov process. When the UE is in state On $S_1$, it will transit to Inactive state $S_2$ if there is a packet within the on-duration timer $T_o$. Otherwise, the UE will transit to state DRX sleep $S_3$. Hence, the state transition probabilities $P_{12}$ and $P_{13}$ are 
\begin{equation}
    P_{12} = \int_0^{T_o} \lambda e^{-\lambda t}\intd{t} = 1-e^{-\lambda T_o},\ \ \ P_{13} = e^{-\lambda T_o}.
\end{equation}

$P_{22}$, $P_{23}$, $P_{31}$ and $P_{13}$ can be calculated similarly. Therefore, we can write the state transition probability matrix as 
\begin{equation}
    P = \begin{bmatrix}
P_{11} & P_{12} & P_{13}\\
P_{21} & P_{22} & P_{23}\\
P_{31} & P_{32} & P_{33}
\end{bmatrix} = 
\begin{bmatrix}
0 & 1-e^{-\lambda T_o} & e^{-\lambda T_o}\\
0 & 1-e^{-\lambda T_i} & e^{-\lambda T_i}\\
e^{-\lambda T_s} & 1-e^{-\lambda T_s} & 0
\end{bmatrix}.
\end{equation}

Then, we denote $\pi_i$ as the steady state probability of state $S_i,\forall i \in \{1,2,3\}$. By applying the balance equations, where $\pi_i = \sum_{j=1}^{3}\pi_j P_{ji}$ and $\sum_{i=1}^3 \pi_i = 1$, we can derive the steady state probabilities as 
\begin{equation}
    \pi_1 = \frac{P_{23}P_{31}}{\Bar{P}},\ \  \pi_2 = \frac{1-P_{13}P_{31}}{\Bar{P}}, \ \ \pi_3 = \frac{P_{23}}{\Bar{P}},
\end{equation}
where $\Bar{P} = 1-P_{13}P_{31}+P_{23}P_{31}+P_{23}$. Then, we will calculate the expected sojourn time in each state. For state $S_1$ and $S_2$, the UE will change the state immediately when a packet arrives. In contrast, the UE in $S_3$ will have a fixed sleep time $T_s$, and the state transition occurs after the UE wakes up. We denote the random variables of sojourn time in state $S_i$ as $W_i,\forall i \in \{1,2,3\}$. For $W_1$, we can calculated $\mathbb{E}[W_1]$ as
\begin{equation}
    \mathbb{E}[W_1] = \int_0^{T_o} \tau \lambda e^{-\lambda \tau} \intd{\tau} + \int_{T_o}^{\infty} T_o \lambda e^{-\lambda \tau} \intd{\tau} = \frac{1-e^{-\lambda T_o}}{\lambda}.
\end{equation}

$\mathbb{E}[W_2]$ can be derived in a similar way. Thus, the expected values of $W_2$ and $W_3$ are:
\begin{equation}
    \mathbb{E}[W_2] = \frac{1-e^{-\lambda T_i}}{\lambda},\ \ \mathbb{E}[W_3] = T_s.
\end{equation}

Finally, the sleep ratio $\gamma$ is the equilibrium probability in $S_3$ of the semi-Markov process. Thus, we can compute $\gamma$ as
\begin{equation}\label{eq:sleep_ratio}
\begin{aligned}
    \gamma &= \frac{\pi_3\mathbb{E}[W_3]}{\sum_{i=1}^3\pi_i\mathbb{E}[W_i]} \\
    &= \frac{P_{23}T_s\lambda}{(1-P_{13}P_{31})P_{22}+P_{23}P_{31}P_{12}+P_{23}T_s\lambda}.   
\end{aligned}
\end{equation}
\subsection{Overview of average AoI calculation}

The TAoI of the status update system is defined as the average age process $\Delta (t)$ in the long term as in \eqref{eq:definition_of_TAoI}. According to the formula proposed in \cite{Kaul2012Realtime}, the TAoI can be obtained by calculating the expected value of the trapezoidal or triangle area $Q_k$ below the AoI function. Moreover, in \cite{Costa2016YkDecomposition}, the authors propose that we can use $Y_k$ and $D_{k-1}$ to compute $\mathbb{E}[Q_k]$, where we have 
\begin{equation}
    \mathbb{E}[Q_k] = \frac{1}{2}\mathbb{E}[Y^2]+\mathbb{E}[Y_k D_{k-1}].
\end{equation}

The start and end points of each $Q_k, k\ge2$ on the time axis are the arrival times at the gNB of the $k-1$st and $k$th received packet, $t_{k-1}$ and $t_k$, respectively. Note that if two packets arrive in the same DRX sleep time, the earlier packet will be discarded, so the UE will not receive all packets successfully. Here, we define $\lambda_e\triangleq \lim_{\tau\to\infty}\frac{N(\tau)}{\tau}$ as the effective arrival rate, where $N(\tau)$ is the number of successfully received packets until time $\tau$. Hereafter, we can rewrite the formula of TAoI $\Delta$ as
\begin{equation}\label{eq:TAoI_formula}
    \Delta = \lambda_e \mathbb{E}[Q_k] = \lambda_e\left(\frac{1}{2}\mathbb{E}[Y_k^2]+\mathbb{E}[Y_k D_{k-1}]\right).
\end{equation}

However, the calculation of TAoI is more complicated, so we will focus on the mathematical derivation of PAoI proposed in \cite{Costa2014peakAge}. PAoI represents the worst case of AoI. It can also characterize the age process well and provide relevant insight. According to \cite{Costa2014peakAge}, the formula of PAoI $\Delta_{(p)}$ can be defined as
\begin{equation}\label{eq:PAoI_formula}
    \Delta_{(p)} = \lim_{\tau\to\infty} \frac{1}{N(\tau)} \left(\sum_{k=1}^{N(\tau)}{D_{k-1} + Y_k}\right) = \mathbb{E}[D_{k-1}] + \mathbb{E}[Y_k].
\end{equation}

Note that from the perspective of the UE, it will only know the generation time and reception time of each successfully received packet. 
Moreover, the expected inter-departure time is the average packet reception interval in the long run, which equals to $\lim_{\tau\to\infty}\frac{\tau}{N(\tau)}$. Therefore, we have

\begin{equation}\label{eq:expected_inter_departure_time}
    \mathbb{E}[Y_k] = \frac{1}{\lambda_e}.
\end{equation}

We must emphasize that $Y_k$ is not an exponential random variable with rate $\lambda_e$. It can be seen from the fact that no packet will be received during DRX sleep, and the packets are not erased randomly. %

\subsection{Effective arrival rate $\lambda_e$}
To compute the effective arrival rate, we denote the event that a packet is not discarded during DRX sleep as $\varphi$. Suppose a packet arrives $\tau$ units of time after DRX sleep has started. If it is not discarded, then there is no packet arrival for the next $T_s-\tau$ units of time. The probability that no packet arrives is a Poisson random variable with rate $\lambda(T_s-\tau)$ and the number of occurrences equals 0, 
then we obtain the probability that event $\varphi$ occurs is
\begin{equation}
\Pr(\varphi)=\int_0^{T_s} e^{-\lambda(T_s-\tau)} \frac{1}{T_s} \;\mathrm{d}\tau=\frac{1-e^{\lambda T_s}}{\lambda T_s}.
\end{equation}

If a packet arrives when a UE is active, the UE can receive the packet successfully with the probability 1. However, if a packet arrives during DRX sleep, the probability that the UE can receive it is $\Pr(\varphi)$. Hence, the effective arrival rate is 
\begin{equation}\label{eq:lambda_e_result}
    \lambda_e = \lambda \left[(1-\gamma) + \gamma\Pr(\varphi)\right]= \lambda\left[1 - \gamma\left(\frac{1-e^{\lambda T_s}}{\lambda T_s}-1\right)\right].
\end{equation}

\subsection{$\mathbb{E}[D_{k-1}]$ and PAoI}
Since the delay time of a received packet is independent of other packets, we omit the subscript $k-1$. Given a packet is successfully received and it arrives during DRX sleep. The expected delay for the packet is $\mathbb{E}[X_\lambda|X_\lambda<T_s]$, where $X_\lambda$ is an exponential random variable with rate $\lambda$.
\begin{equation}
\begin{aligned}\label{eq:expected_delay_for_pkt_in_sleep}
        \mathbb{E}[X_\lambda|X_\lambda<T_s] &= \frac{1}{1-e^{-\lambda T_s}}\int_0^{T_s}\lambda t e^{-\lambda t}\;\mathrm{d}t\\
    &= \frac{1}{\lambda}-\frac{T_s e^{-\lambda T_s}}{1-e^{-\lambda T_s}}.
\end{aligned}
\end{equation}

There is no delay when the packets are received while the UE is active. Therefore, we multiply the delay calculated in \eqref{eq:expected_delay_for_pkt_in_sleep} by the percentage of packets that arrive during DRX sleep among all successfully received packets. In this way, we can write the total delay as
\begin{equation}\label{eq:delay_result}
\begin{aligned}
            \mathbb{E}[D] &= \frac{\gamma\Pr(\varphi)}{1-\gamma+\gamma\Pr(\varphi)}\mathbb{E}[X_\lambda|X_\lambda<T_s]\\
            &=\frac{\gamma(1-e^{-\lambda T_s})}{(1 - \gamma)\lambda T_s + \gamma (1-e^{-\lambda T_s})}\left(\frac{1}{\lambda}-\frac{T_se^{-\lambda T_s}}{1-e^{-\lambda T_s}}\right).
\end{aligned}
\end{equation}

Finally, by combining \eqref{eq:expected_inter_departure_time}, \eqref{eq:lambda_e_result} and \eqref{eq:delay_result}, we can obtain the closed-form solution of PAoI:
\begin{equation}\label{eq:PAoI_result}
    \Delta_{(p)} = \frac{\frac{1}{\lambda}+\gamma(\frac{1}{\lambda}-\frac{T_s e^{-\lambda T_s}}{1-e^{-\lambda T_s}})}{\frac{1-\gamma}{(1-e^{-\lambda T_s})}\lambda T_s+\gamma}.
\end{equation}

\section{Two-phase DRX service policy}\label{sec:service_policy}
A two-phase DRX service policy is proposed to minimize average AoI performance by appropriately calibrating the DRX mechanism's parameters. By inspecting the results in \eqref{eq:sleep_ratio} and \eqref{eq:PAoI_result}, we observe that:
\begin{enumerate}
    \item The system performance is highly related to the sleep time $T_s$. A lower $T_s$ implies a lower average AoI. This is because a short UE sleep time can significantly reduce packet delay when the UE is sleeping. Therefore, setting $T_s$ to the minimum sleep time $T_{s}^{\min}$ and making the UE switch between sleep and active states more frequently may result in a lower average AoI.
    \item The lower inactivity timer $T_i$ means a better average AoI under a given sleep ratio. Because the original design of the inactivity timer considers the assumption that when a packet is received, further packets will likely arrive within a short time. This assumption is reasonable for applications such as web browsing and can handle busty traffic properly. However, for the Poisson arrival model considered in this study, the UE prolongs the active time unnecessarily. If the UE sets the inactivity timer to 0 ms, the UE starts sleeping after the reception of the first packet in the on-duration.
\end{enumerate}

As a result, by fixing the sleep time to the minimum possible value $T_{s}^{\min}$ and inactivity timer $T_i = 0$, we can derive the corresponding proposed on-duration timer $T_o^\star$ under a given sleep ratio $\gamma$ by \eqref{eq:sleep_ratio}:
\begin{equation}\label{eq:proposed_on_duration}
    T_o^\star = \frac{1}{\lambda}\log\left(\frac{\gamma}{\gamma-(1-\gamma)T_s^{\min} \lambda e^{T_{s}^{\min}\lambda}}\right).
\end{equation}

The proposed on-duration timer $T_o^\star$ should always be greater or equal to zero. By organizing \eqref{eq:proposed_on_duration}, we find that there exists a minimum value of sleep ratio $\gamma^\star$ to guarantee $T_o^\star \ge 0$:
\begin{equation}\label{eq:gamma_threshold}
    \gamma^\star = \frac{\lambda T_{s}^{\min}e^{\lambda T_{s}^{\min}}}{\lambda T_{s}^{\min}e^{\lambda T_{s}^{\min}}+1}.
\end{equation}

If the required sleep ratio is lower than $\gamma^\star$, even if $T_o$ is set to infinity and $T_s$ is set to the minimum value, it is not possible to make the ratio of the UE in sleep smaller than the required sleep ratio $\gamma$. Therefore, we inevitably need to set $T_i$ larger than 0. By setting $T_o\to\infty$ and $T_s = T_{s}^{\min}$, the corresponding proposed inactivity timer $T_i^\star$ calculated by \eqref{eq:sleep_ratio} is
\begin{equation}\label{eq:proposed_inactive}
    T_i^\star = \frac{1}{\lambda}\log\left(1+\frac{1-\gamma}{\gamma}T_{s}^{\min}\lambda-e^{-\lambda T_{s}^{\min}}\right).
\end{equation}

Consequently, by combining \eqref{eq:proposed_on_duration} to \eqref{eq:proposed_inactive}, we propose an intuitive and easy-to-implement two-phase DRX service policy $\pi(\gamma)$, which can give the best parameters setting of the DRX mechanism to minimize average AoI as the following:
\begin{equation}
\begin{aligned}
    \pi(\gamma) =     
            \begin{cases}
            (T_o^\star, 0, T_{s}^{\min}) & \gamma > \gamma^\star\\
            (\infty, T_i^\star, T_{s}^{\min}) & \text{otherwise.}
            \end{cases}
    \end{aligned}
\end{equation}

\section{Simulations and AoI performance analysis}\label{sec:simulations}
In this section, we perform the simulation to justify that PAoI and TAoI have the same variation trend with DRX parameters settings. Moreover, the simulations also show the proposed two-phase DRX service policy can be an excellent strategy to minimize TAoI under a given sleep ratio.
\begin{figure*}
     \centering
     \begin{subfigure}[b]{0.3\textwidth}
         \centering
         \includegraphics[width=\textwidth]{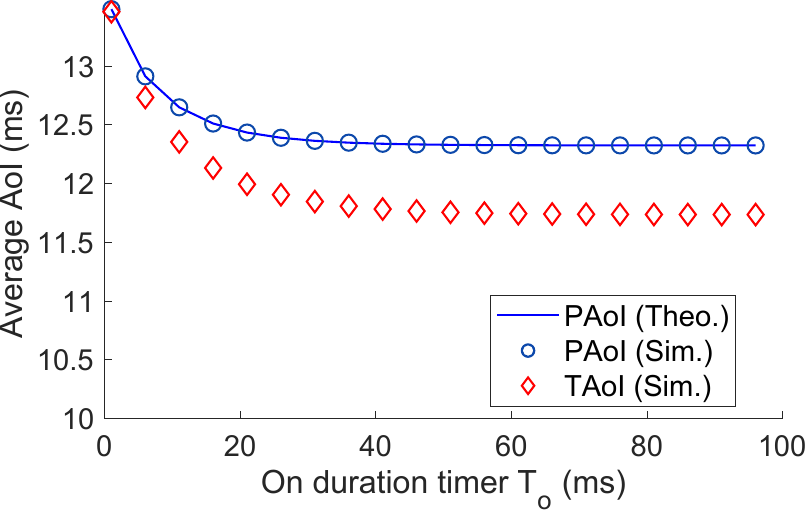}
         \caption{$T_i = 10$ ms, $T_s=10$ ms}
         \label{fig:fix_To}
     \end{subfigure}
     \hfill
     \begin{subfigure}[b]{0.3\textwidth}
         \centering
         \includegraphics[width=\textwidth]{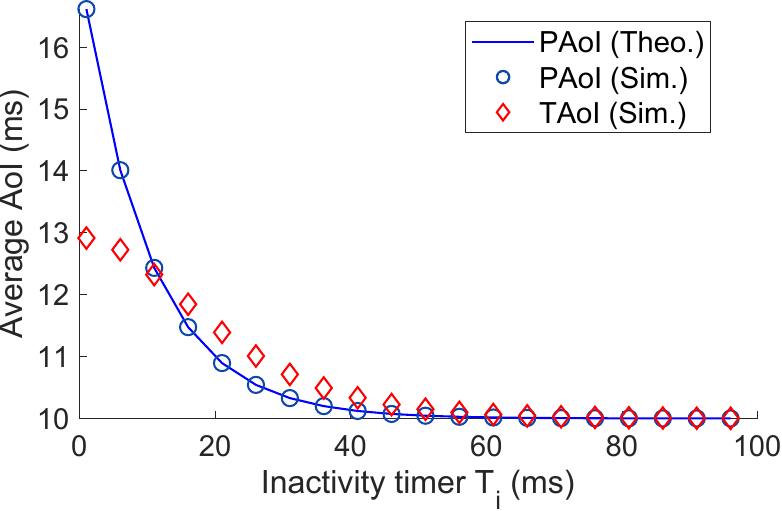}
         \caption{$T_o = 10$ ms, $T_s=10$ ms}
         \label{fig:fix_Ti}
     \end{subfigure}
     \hfill
     \begin{subfigure}[b]{0.3\textwidth}
         \centering
         \includegraphics[width=\textwidth]{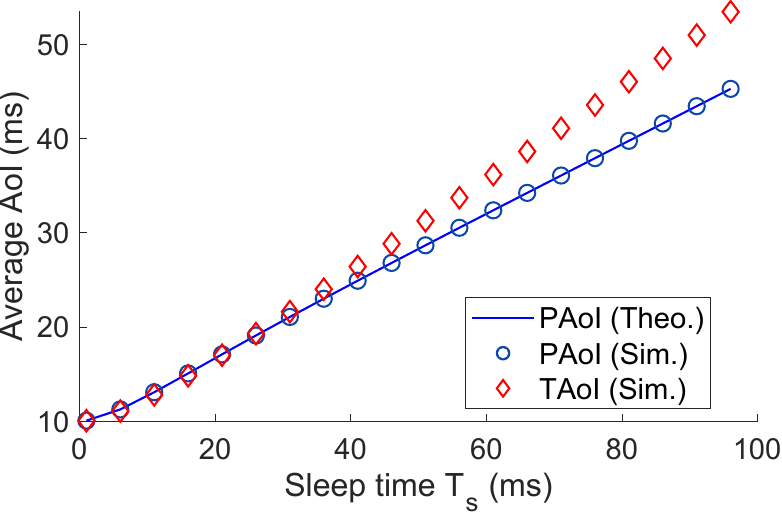}
         \caption{$T_o = 10$ ms, $T_i=10$ ms}
         \label{fig:fix_Ts}
     \end{subfigure}
        \caption{PAoI and TAoI with respect to different DRX parameter settings}
        \label{fig:PAoI_TAoI}
\end{figure*}

\begin{figure*}
    \centering
    \includegraphics[width=\linewidth]{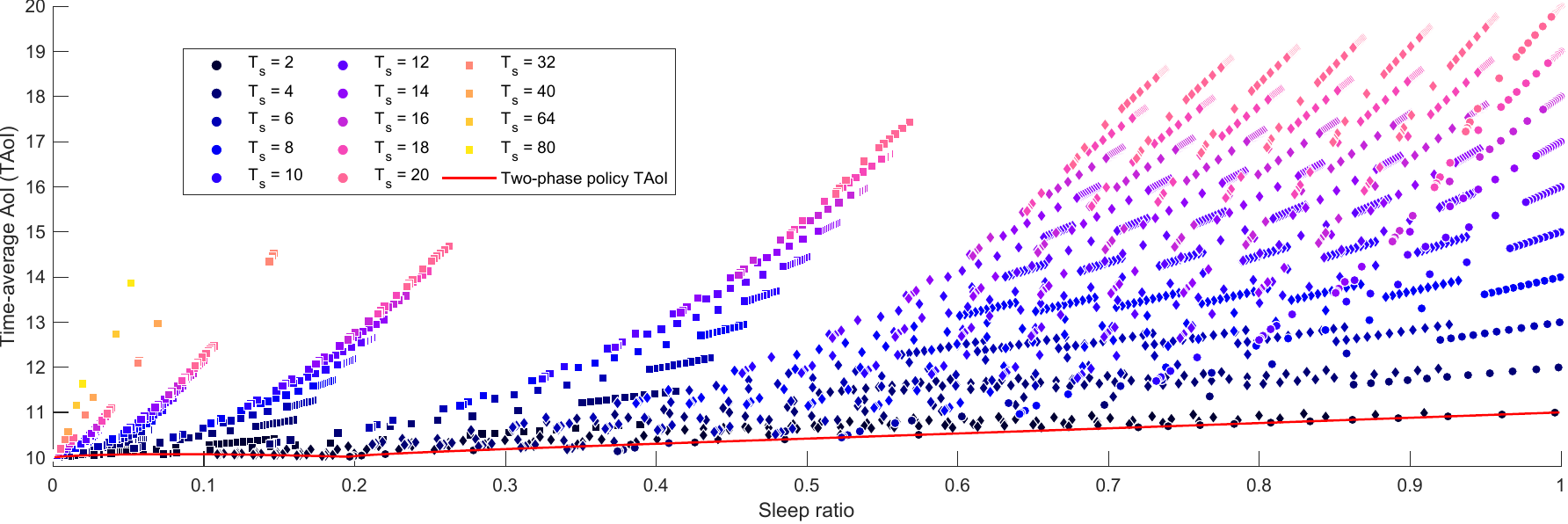}
    \caption{TAoI and sleep ratio with different DRX parameter settings. The points are colored by their DRX sleep time, and the shape of the data points indicates their inactivity timer. The circle points represent $T_i = 0$, the diamond points mean $0<T_i<10$, and the square points have $10\le T_i <100$. The red line is the TAoI obtained by the two-phase DRX service policy $\pi(\gamma)$.}
    \label{fig:scatter}
\end{figure*}
\subsection{Evaluation Methodology}
The theoretical value of PAoI is obtained by \eqref{eq:PAoI_result}, and the simulation value of PAoI and TAoI can be calculated via Monte-Carlo simulation by the formula \eqref{eq:TAoI_formula} and \eqref{eq:PAoI_formula}, respectively. By the definition in \eqref{eq:definition_of_TAoI} and to get more stable results, every sample point of TAoI and PAoI has the total simulation time $T=10^9$ ms and is averaged over $10000$ runs. All the simulations are conducted in MATLAB. The arrival rate is $\lambda = 0.1$ packet/ms for the following simulations. 

We can obtain the lower bound of TAoI in the system by \eqref{eq:TAoI_formula}. Assuming the DRX is not applied, UE will successfully receive every packet without delay. Hence, the random variable of the inter-departure time is simply an exponential random variable with the rate $\lambda$, and the term $\mathbb{E}[Y_k D_{k-1}]$ will equal 0. As a result, the lower bound of TAoI $\Delta_{LB}$ is
\begin{equation}
    \Delta_{LB} = \lambda\left(\frac{1}{2} \cdot \frac{2}{\lambda^2} + 0\right) = \frac{1}{\lambda}.
\end{equation}

Since we have $\lambda = 0.1$ packet/ms, $\Delta_{LB} = 10$ ms is the minimum achievable average AoI in the following simulations.

\subsection{PAoI and TAoI comparisons}

We change the values of the different DRX parameters setting to observe the variation of PAoI and TAoI.
\cref{fig:PAoI_TAoI} shows that the derived theoretical PAoI and the simulated TAoI have a similar trend no matter how we adjust the parameters. Therefore, it verifies that PAoI can be an alternative to calculate the average AoI efficiently and understand the trend of TAoI. 

\cref{fig:fix_To} implies that even if we adjust $T_o$ to a very high value, average AoI converges to a stable value but does not reach the lower bound $\Delta_{LB}$. In contrast, \cref{fig:fix_Ti} shows as $T_i$ increases, average AoI reaches $\Delta_{LB}$. The reason is that the inactivity timer restarts whenever the UE receives a packet during the inactive state, so it is likely that the UE stays in an inactive state for a long time when $T_i$ is large. However, the trade-off of increasing $T_i$ will be a meager sleep ratio.
Finally, the y-axis of the \cref{fig:fix_Ts} shows that increasing the sleep time $T_s$ of the UE significantly affects the average AoI. 

\subsection{TAoI and sleep ratio performance analysis}\label{subsec:sim_2}
We numerically evaluate the TAoI and sleep ratio with different DRX parameter settings in \cref{fig:scatter}. Moreover, we also simulate the performance of the proposed two-phase DRX service policy. We list all possible DRX parameters of this simulation in \cref{tab:DRX_parameters}. The values of the parameters are referenced from \cite{3gpp.38.331} and \cite{Koc2014DRXmodel}. Each data point in \cref{fig:scatter} is derived from a possible combination of the DRX parameters in \cref{tab:DRX_parameters}, and the location of the points are determined by their TAoI and sleep ratio simulation result. \cref{fig:scatter} will not show the data points with TAoI larger than $20$ ms. We have $T_{s}^{\min} = 2$ ms in this simulation, and thus $\gamma^\star = 0.1963$ by \eqref{eq:gamma_threshold}.

As shown in fig. \ref{fig:scatter}, DRX sleep time will heavily affect TAoI even when the sleep ratio is fixed. For a given sleep ratio, a lower DRX sleep time means a better TAoI. Additionally, when $\gamma>\gamma^\star$, the points with lower TAoI are mostly circular, i.e., the points with $T_i=0$, meaning that not activating the inactivity timer can minimize the average AoI. These two results support our observations in \ref{sec:service_policy} for average AoI and DRX parameters. Thus, our proposed two-phase DRX service policy, the red line in fig. \ref{fig:scatter}, is quite close to the lower bound of the average AoI $\Delta_{LB}$ even when $\gamma>\gamma^\star$, and outperforms most of the data points with other DRX parameters. However, for sleep ratio $\gamma$ between 0.3 and 0.4, a few points with $T_s=4$ have better TAoI than those with $T_s=T_{s}^{\min}$. It suggests that a slight increase in sleep time may improve the system performance when $\gamma$ is close to $\gamma^\star$. But in any case, using a lower $T_s$ is a reliable strategy to reduce the overall TAoI.

\begin{table}
    \centering
    \smallskip
    \caption{DRX parameters values in simulation \ref{subsec:sim_2}}
    \begin{tabular}{cp{.62\columnwidth}}
    \toprule
        Parameter & Values (ms)  \\
    \midrule
       on-duration timer $T_o$ & 0.01, 0.1, 0.2, 0.3, 0.4, 0.5, 0.6, 0.7, 0.8, 0.9, 1, 2, 3, 4, 5, 6, 10, 20, 30, 40, 50, 60\\
       inactivity timer $T_i$ &0, 1, 2, 3, 4, 5, 6, 10, 20, 30, 40, 50, 60\\
       DRX sleep time $T_s$ &2, 4, 6, 8, 10, 12, 14, 16, 18, 20, 32, 40, 64, 80\\
    \bottomrule
    \end{tabular}
    \label{tab:DRX_parameters}
\end{table}

\section{Conclusion}\label{sec:conclusion}
This study analyzes the impact on data freshness when a UE uses the DRX mechanism, as measured by AoI. We derive a closed-form solution for the sleep ratio and PAoI. In addition, we propose a practical strategy to adjust DRX parameters. Based on simulations, we verify that the more easily calculated PAoI shows a similar trend to TAoI. The UE should reduce the DRX sleep time in most cases to obtain better AoI. It should also avoid turning on the inactivity timer to save power. The results confirm our proposed DRX parameter strategy achieves TAoI close to the theoretical AoI minimum.

\bibliographystyle{IEEEtran}
\bibliography{references}
\vspace{12pt}

\end{document}